# Blistering Failure of Elastic Coatings with Applications to Corrosion Resistance

Surya Effendy[a], Tingtao Zhou[b], Henry Eichman[c], Michael Petr[d], and Martin Z. Bazant[a,e,1]

[a]Department of Chemical Engineering, Massachusetts Institute of Technology, 77 Massachusetts Avenue, Cambridge, MA 02138, USA
[b]California Institute of Technology, Division of Engineering and Applied Sciences, 1200 E. California Blvd., Pasadena, CA91125, USA [c]Dow Coating Materials, 400 Arcola Road, Collegeville, PA19426, USA
[d]Dow Wire and Cable, 400 Arcola Road, Collegeville, PA19426, USA [e]Department of Mathematics, Massachusetts Institute of Technology, 77 Massachusetts Avenue, Cambridge, MA 02138, USA

## Abstract

A variety of polymeric surfaces, such as anti-corrosion coatings and polymer-modified asphalts, are prone to blistering when exposed to moisture and air. As water and oxygen diffuse through the material, dissolved species are produced, which generate osmotic pressure that deforms and debonds the coating. These mechanisms are experimentally well-supported; however, comprehensive macroscopic models capable of predicting the formation osmotic blisters, without extensive data-fitting, is scant. Here, we develop a general mathematical theory of blistering and apply it to the failure of anti-corrosion coatings on carbon steel. The model is able to predict the irreversible, nonlinear blister growth dynamics, which eventually reaches a stable state, ruptures, or undergoes runaway delamination, depending on the mechanical and adhesion properties of the coating. For runaway delamination, the theory predicts a critical delamination length, beyond which unstable corrosion-driven growth occurs. The model is able to fit multiple sets of blister growth data with no fitting parameters. Corrosion experiments are also performed to observe undercoat rusting on carbon steel, which yielded trends comparable with model predictions. The theory is used to define three dimensionless numbers which can be used for engineering design of elastic coatings capable of resisting visible deformation, rupture, and delamination.

## 1. Introduction

The global cost of corrosion in 2019 is approximately 3 trillion USD, of which between 15% to 35% is preventable [1, 2]. For this reason, there have been concerted political efforts for a more rigorous implementation of corrosion prevention technology in civilian, military, and industrial contexts [3, 4]. Such technology includes, but is not limited to, manipulation of surface [5, 6] and bulk geometry [7, 8, 9], design of corrosion-resistant alloys [10, 9], elimination of oxidative environments, application of corrosion inhibitors [11, 12], anodic protection [13], cathodic protection via sacrificial anode [14] or impressed current systems [15], and application of organic [16, 17] and inorganic coatings [18, 19].

---

[1] Corresponding author: bazant@mit.edu

Of the existing corrosion prevention technology, aside from metallurgical modifications, protective coatings remain the most widely used. Various failure mechanisms have been discussed in the literature, ranging from loss of adhesion due to particulate inclusions [6] or chemical reactions [20, 21, 22], salt or solvent entrapment [23, 24, 25, 26, 27], microstructural defects [24, 16, 6, 28], solvent absorption [23], decreased plasticity due to polymer degradation [29, 30] or loss of plasticizers [31, 32], and miscellaneous environmental stresses (e.g., accidental scratches) [33, 34]. These mechanisms are well-supported by experimental and field observations [25, 35]; however, continuum models capable of predicting failure of protective coatings, without extensive data-fitting, is scarce. This limits the pace of coating development, which is presently reliant on time- and resource-intensive field tests [36, 37, 38].

In the present work, we focus on osmotic blistering as the primitive form of the failure mechanisms listed above. In section 2, we develop a general mathematical model capable of predicting the formation of blisters at a macroscopic scale, and simplify the formulation by assuming an axisymmetric blister overlaying carbon steel, a situation commonly encountered in field applications. In section 3, we introduce the experimental data sets used to validate the model and the methodology used to analyze raw data sets taken from the literature. In section 4, we show that the model is able to fit the experimental data sets, validating the hypothesis of an irreversible, nonlinear blister growth rate, and more generally, the growth and deformation of blisters under the influence of osmotic pressure. The model predicts a long-term behavior characterized by a stable, ruptured, or delaminated blister and outlines the initiation, propagation, and termination steps of the blistering process. We then use the theory to define three dimensionless numbers, each classified according to the source of osmotic pressure, which can be used for engineering design of polymeric coatings capable of resisting visible deformation, rupture, and delamination. In section 5, we discuss the limitations of the model, as well the various other contexts within which the model is applicable. In section 6, we summarize the main findings of the work.

## 2. Mathematical Model

Osmotic blisters are formed due to the tendency for water to move from regions of low solute activity to regions of high solute activity. When this transfer occurs across a semi-permeable membrane, i.e., a membrane permeable to water but not to solutes, pockets of soluble materials hydrate and expand, exerting an osmotic pressure on the surrounding matrix. If these pockets are trapped between a coating and a metal surface, the resulting osmotic pressure can push the two materials apart, leading to coating failure.

The entrapped water-soluble components are often salts that are deliberately introduced to modify the properties of the coating or transferred to the metal surface during the coating application process [26, 27]. Small organic compounds have also been considered as possible water-soluble components, especially for poorly-formulated solvent-based paints [23]. Blisters may then develop upon exposure to rain, fog, or other domestic and industrial water sources (e.g., bathroom and kitchen appliances, water tanks).

To illuminate the osmotic blistering mechanism, we will begin by presenting the general form of the model and then provide a more practical axisymmetric formulation.

## 2.1. Full Model

We treat the coated metal surface as an infinite two-dimensional surface with Cartesian coordinates $x$ and $y$ immersed in an infinite reservoir of water. The *adhesion state* is characterized by a binary variable $\xi$, illustrated in figure 1(a):

$$\xi(x,y) = \begin{cases} 0 \text{ if the coating is debonded from the metal at } (x,y) \\ 1 \text{ otherwise} \end{cases} \tag{1}$$

With the implicit assumption that the debonded area is finite, the above formulation is equivalent to the set of perimeters $\{\Omega_i\}$, with each perimeter $\Omega_i$ describing a distinct contiguous debonded area. With no loss of generality, we will assume a single contiguous debonded area with perimeter $\Omega$ (see figure 1(b)) satisfying the parametric equation $\omega(s) = (\omega_x(s), \omega_y(s))$.

This perimeter is associated with a contiguous volume $V$ containing blister fluid at pressure $\Pi$. Assuming elastic, isotropic, and sufficiently linear material properties, the coating deforms freely:

$$\Pi = \gamma \nabla \cdot \hat{\boldsymbol{n}} \tag{2}$$

Here $\gamma$ is the effective surface tension of the coating and $\hat{\boldsymbol{n}}$ the unit vector normal to the deformed surface, illustrated in figure 1(c). Equation 2 is solved with the boundary condition:

$$h(\omega_x(s), \omega_y(s)) = 0 \tag{3}$$

Here $h(\cdot, \cdot)$ is the blister height at a given coordinate.

Most coatings of practical interest are viscoelastic. However, their viscoelastic relaxation times ($\sim 1$ hour [39, 40]) tend to be much shorter than the timescale associated with the formation of blisters and the attendant coating failure ($\sim$1 day to 1 year). We thus treat the coating as an elastic material, which can be modelled as a Mooney-Rivlin solid [41, 42]:

$$\gamma = s_1 l(t = 0) \left(1 - \left[\frac{A_{\text{base}}}{A}\right]^3\right)\left(1 - \frac{s_1}{s_{-1}}\frac{A}{A_{\text{base}}}\right) \tag{4}$$

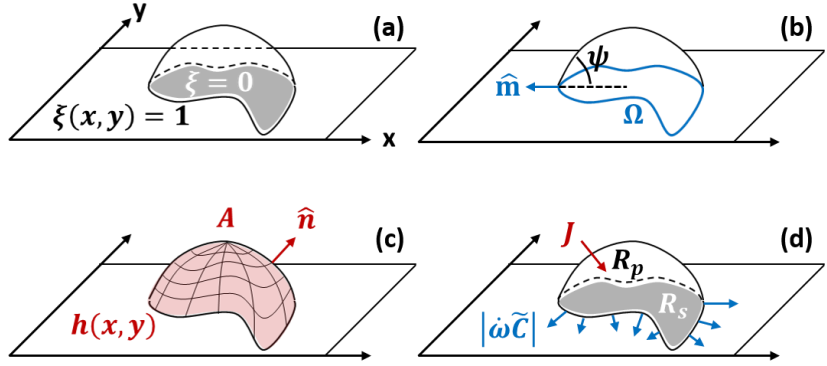

Figure 1: (a) Isometric view of a blister. Adhesion state is characterized by a binary variable $\xi(x,y)$, which is 0 if the coating is debonded at $(x,y)$ and 1 otherwise. (b) For finite blisters, adhesion state is dual to the perimeter $\Omega$ characterized by the unit normal vector $\hat{\boldsymbol{m}}$. The local contact angle $\psi$ is evaluated in the direction of $\hat{\boldsymbol{m}}$. (c) Pressure differential (osmotic or otherwise) results in a surface curvature described by the height function $h(x,y)$ with total surface area $A$. The surface curvature is characterized by the unit normal vector $\hat{\boldsymbol{n}}$. (d) Our realization of the model considers the effect of heterogeneous reaction rate at the exposed surface $R_s$, homogeneous precipitation rate in the bulk volume $R_p$, diffusion through the polymeric coating $J$, and dissolution through the moving perimeter $|\dot{\boldsymbol{\omega}}\tilde{C}|$.

Here $s_1$ and $s_{-1}$ are parameters which depend upon the coating under consideration, $l(t = 0)$ is the initial / unstretched thickness of the coating, and $A$ and $A_{\text{base}}$ are the stretched and unstretched areas of the coating associated with $\Omega$, respectively. Equation 4 is only valid up to elongation at failure :

$$\sqrt{\frac{A}{A_{\text{base}}}} \leq 1 + \epsilon \tag{5}$$

Beyond this point, the blister breaks. Note that equation 4 neglects bending moment, which can be included to obtain a more accurate understanding of coating deformation in cases whereby the blister size is of the same order as the coating thickness.

At the perimeter, the interface between the coating and the metal experiences an *adhesion stress*, which causes the coating to debond from the metal:

$$\frac{d\boldsymbol{\omega}}{dt} \equiv \dot{\boldsymbol{\omega}} = k_{\text{ad}}\hat{\boldsymbol{m}}(\sigma_{\text{ad}} - \sigma_{\text{ad}}^*)H(\sigma_{\text{ad}} - \sigma_{\text{ad}}^*) \tag{6}$$

Here $k_{\text{ad}}$ is the mobility of the interface, $\hat{\boldsymbol{m}}$ is the unit vector normal to the perimeter at $\omega$ on the $(x,y)$ plane (see figure 1(b)), $\sigma_{\text{ad}}$ is the adhesion stress, $\sigma_{\text{ad}}^*$ is the critical adhesion stress, and $H(\cdot)$ is the Heaviside function. The critical adhesion stress $\sigma_{\text{ad}}^*$ is an interfacial property which describes stickiness. The adhesion stress describes the magnitude of debonding force at the perimeter:

$$\sigma_{\text{ad}} = \frac{\gamma(\psi)}{\lambda} \tag{7}$$

Here $\psi$ is the local contact angle, which is evaluated from:

$$\tan(\psi) = -\nabla h \cdot \hat{\boldsymbol{m}} \tag{8}$$

and $\lambda$ is a characteristic length over which elastic forces are spread. For an elastic coating with no bending moment, and assuming large blister size relative to coating thickness, this characteristic length is approximately the projected thickness of the coating (visualized in figure 2(c,d) for an axisymmetric blister) [25], which is obtained by solving equation 2.

The pressure $\Pi$ can be effected by inertial, viscous, or body forces, but the present work focuses on osmotic pressure. We assume quasi-steady solute distribution within the blister, which holds for most cases of practical interest:

$$\text{Bi} = \frac{D_{\text{coat}}\,L_{\text{fluid}}}{D_{\text{fluid}}\,L_{\text{coat}}} \sim 0.001$$

Here Bi is the Biot number, which describes the mass-transfer resistance of the blister fluid relative to the coating, $D_{\text{coat}}$ the mass diffusivity of the soluble components in the coating, $D_{\text{fluid}}$ the mass diffusivity of the soluble components in the blister fluid, $L_{\text{coat}}$ the thickness of the coating, and $L_{\text{fluid}}$ the size of the blister. The osmotic pressure can then be calculated from the van't Hoff theory:

$$\Pi = \sum_i \left[\frac{\gamma_i n_i}{V} - \gamma_{i,\text{out}} C_{i,\text{out}}\right] k_B T \tag{9}$$

Here $\gamma_i$ is the activity coefficient of species $i$, $n$ the amount of soluble species, $C_{\text{out}}$ the concentration in the bulk electrolyte, $k_B$ the Boltzmann constant, and $T$ the temperature.

The amount of soluble species $n_i$ is tracked using mass balance, as illustrated in figure 1(d):

$$\frac{dn_i}{dt} = J_i A + \nu_i R_s A_{\text{base}} + \eta_i R_p V + \oint |\dot{\boldsymbol{\omega}}\tilde{C}_i| d\Omega \tag{10}$$

The four terms on the right-hand side of equation 10 correspond to flux through the coating $J_i(V, n_i)$, heterogeneous reaction rate at exposed metallic surfaces $R_s(V, n_i)$ with stoichiometric coefficients $\nu_i$, precipitation reaction rate in the blister fluid $R_p(V, n_i)$ with stoichiometric coefficients $\eta_i$, and flux through the moving perimeter of the blister, where $\bar{C}(x, y)$ is the surface concentration of soluble species along the metal / coating interface. The forms of $J_i$, $R_s$, $R_p$, and $\bar{C}$ vary according to the choice of metal, coating, and coating application.

### 2.2. Axisymmetric Blister

As illustrated in figure 2(a,b), most blisters of practical interest are approximately radially symmetric, and can be described by the contact angle $\psi$ and the base radius $r$. With this geometry, equations 2, 4, and 5, which describe coating deformation and failure, are simplified to:

$$\Pi = \frac{2\gamma \sin \psi}{r} \tag{11}$$

$$\gamma = s_1 l(t=0) \left[1 - \left(\frac{\sin^2 \psi}{2[1-\cos \psi]}\right)^3\right] \left[1 - \frac{s_1}{s_{-1}} \frac{2(1-\cos \psi)}{\sin^2 \psi}\right] \tag{12}$$

$$\frac{2(1-\cos \psi)}{\sin^2 \psi} \leq 1 - \epsilon \tag{13}$$

The debonding rate, i.e., equation 6 simplifies to:

$$\frac{dr}{dt} = k_{\text{ad}}(\sigma_{\text{ad}} - \sigma_{\text{ad}}^*) H(\sigma_{\text{ad}} - \sigma_{\text{ad}}^*) \tag{14}$$

To reiterate, equations 11 and 14 describe the balance between elastic and osmotic forces, and the tendency for interfacial bonds to break under stress, respectively. These are illustrated for the axisymmetric case in figure 2(c,d).

Evaluation of adhesion stress requires additional knowledge of material compressibility. Most coatings of practical interest are designed to be rubbery at room temperature and thus approximately incompressible [43], leading to the expression:

$$\sigma_{\text{ad}} = \frac{\gamma}{l(t=0)} \frac{2(1-\cos \psi)}{\sin^2 \psi} \tag{15}$$

The osmotic pressure is evaluated assuming an ideal solution:

$$\Pi = \sum_i \left[\frac{n_i}{V} - C_{i,\text{out}}\right] k_B T \tag{16}$$

The mass balance, i.e., equation 10 is evaluated assuming that all soluble contaminants are initially concentrated at a spot:

$$\frac{dn_i}{dt} = J_i A + \nu_i R_s A_{\text{base}} + \eta_i R_p V \tag{17}$$

Mass transport in the coating is assumed to be Fickian and quasisteady with respect to $C$, leading to the simple expression:

$$J_i = D_i \frac{C_{i,\text{out}} - C_i}{l(t=0)} \frac{2(1 - \cos \psi)}{\sin^2 \psi}$$

(18)

Here $C_{i,\text{out}}$ is the concentration of species $i$ in the bulk electrolyte.

Equations 11 through 18 complete the mechanical and transport components of the axisymmetric model, which is general with respect to the chemistry of the corrosion reaction. We now turn our attention to a specific realization of the corrosion chemistry.

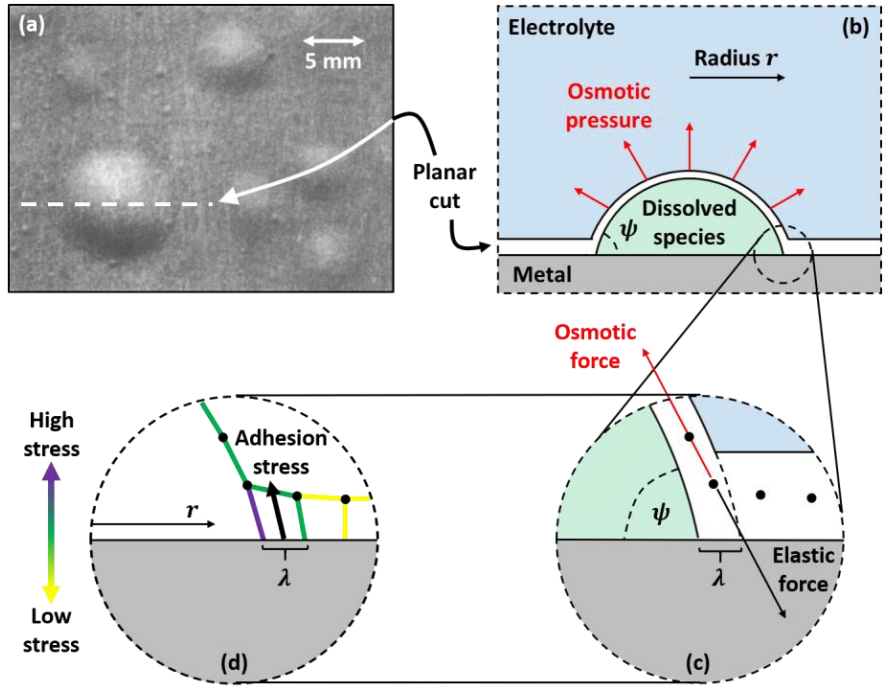

Figure 2: (a)Light microscopy image of circular blisters formed on epoxy-coated aluminum, taken from [27]. Further evidence of circular blisters formed on polymer-coated metals can be found in [23, 44, 24, 6]. Schematic of the blister along the planar cut shown in (a). Osmotic pressure generated by corrosion products and soluble inclusions stretches the coating into a sphere segment with contact angle $\psi$ and base radius $r$ assuming isotropic mechanical properties. (c) A close-up view of the perimeter of the coating shows the balance between osmotic force, obtained by integrating osmotic pressure across the surface of the blister, and the elastic restoring force, which occurs due to polymer stretching. (d) The elastic restoring force incurs a stress on the polymer near the perimeter of the blister. This is the adhesion stress, which must exceed the yield strength of the polymer-metal interface for the blister to grow. The elastic restoring force is spread over a characteristic length scale $\lambda$, which is assumed to be the projected thickness of the coating.

### 2.3. Corrosion Chemistry

To simplify the system under consideration, we will focus on polymer-coated iron, whose corrosion rate is known to be limited by the transport of oxygen [45]:

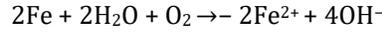

$$R_s = -\frac{D_{O_2}}{\nu_{O_2}} \frac{C_{O_2,\text{out}} - C_{O_2}}{l(t=0)} \left[ \frac{2(1-\cos\psi)}{\sin^2\psi} \right]^2 \tag{19}$$

Our in-house experiments suggest that the concentration of oxygen within the blister is sufficiently low such that $Fe^{2+}$ is preferentially formed over $Fe^{3+}$, up to the point where visible blisters form. This corrosion reaction proceeds according to the stoichiometry:

$$2Fe + 2H_2O + O_2 \rightarrow 2Fe^{2+} + 4OH^-$$

Thus $\nu_{O_2} = -1$, $\nu_{Fe^{2+}} = 2$, and $\nu_{OH^-} = 4$. A more complete discussion of these in-house experiments can be found in section 4.2.

The corrosion reaction results in accumulation of soluble species, which eventually exceeds its maximum solubility:

$$Fe^{2+} + 2OH^- \rightarrow Fe(OH)_2$$

In most cases of practical interest, iron corrodes within an electrolyte containing some amount of aggressive anions (i.e., anions of strong acids), which can greatly accelerate the corrosion process [46]. To improve the utility of the model, we will consider an electrolyte containing NaCl, which modifies the precipitation equilibrium by associating with $Fe^{2+}$ ions:

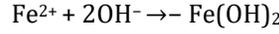

$$Fe^{2+} + Cl^- \leftrightarrow FeCl^+ \tag{20}$$

For simplicity, we will treat $Fe^{2+}$ and $FeCl^+$ as identical as far as the mass balance, i.e., equation 17 is concerned. The effect of equation 20 is thus limited to modifying the solubility limit of $Fe(OH)_2$:

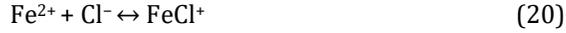

$$S^*_{Fe(OH)2} = S_{Fe(OH)2}(1 + K_{FeCl+}C_{Cl-}) \tag{21}$$

Here $S^*_{Fe(OH)2}$ and $S_{Fe(OH)2}$ are the solubility limits in the presence and absence of chloride ions, respectively, and $K_{FeCl+}$ is the equilibrium constant of equation 20.

In the present work, we will assume a linear non-equilibrium expression for precipitation kinetics, driven by chemical potential difference [47]:

$$R_p = k_p\left(-\eta_{Fe2+}k_BT\ln C_{Fe2+} - \eta_{OH-}k_BT\ln C_{OH-} - \mu^{\ominus}\right) \tag{22}$$

Here $k_p$ is the rate constant of the precipitation reaction and $\mu^{\ominus}$ is the chemical potential at the solubility limit. The latter is evaluated from:

$$\mu^\ominus = k_B T \ln\left[S^*_{Fe(OH)2}\right] - \left[\frac{\beta_p}{c_{Fe(OH)2}}\right]^2 \tag{23}$$

The second term on the right-hand side of equation 23 is a chemical potential barrier which prevents rust dissolution in the absence of rust. Equation 22 is a simplification of true rust precipitation kinetics, which involves slow condensation of a gel-like polymer phase followed by a relatively fast dehydration / crystal growth step [48]. The true kinetics is further complicated by the presence of aggressive anions, whose impact is still not fully understood [46]. However, the timescale for rust precipitation (~minutes [48]) is much smaller than the timescale for blister formation. In that regard, the exact form of the precipitation rate is not expected to be important, beyond the fact that $k_p$ is large.

*2.4. Nondimensionalization*

We set the length scale of the problem to the initial thickness of the coating:

$$l_{\text{ref}} = l(t = 0) \tag{24}$$

The concentration scale is set by the saturation limit of $Fe(OH)_2$:

$$C_{\text{ref}} = \left[S_{Fe(OH)2}\left(-\frac{z_{Fe2+}}{z_{OH-}}\right)^{\eta_{OH-}}(1 + K_{FeCl+}C_{Cl-})\right]^{-\frac{1}{\eta_{Fe2+}+\eta_{OH-}}} \tag{25}$$

The time scale is set by the transport of oxygen through the coating:

$$t_{\text{ref}} = \frac{l(t = 0)^2}{D_{O_2}} \tag{26}$$

The resulting nondimensionalized equations are summarized in the supplementary information.

*2.5. Numerical Implementation*

The model presented in table 2 is solved on MATLAB R2020a using the command ode15s. The model combines differential and algebraic equations, which necessitates the *mass* option. The code is sped up using the *vectorized* option. The absolute and relative tolerances are set to $10^{-8}$ and $10^{-6}$, respectively. Each simulation takes less than a second to complete.

## 3. Experimental Data and Analysis

In this section, we briefly describe the key assumptions and consequences of the model, and describe the experiments which have been performed to validate those assumptions and the model in general. The data in sections 3.1 and 3.2 is taken from the literature [25, 35], while the data in section 3.3 has been generated specifically for the present work.

### 3.1. Irreversible, Nonlinear Growth Rate

Equation 6 predicts a linear relationship between growth rate and adhesion stress, but only if the adhesion stress exceeds a critical value [49]. The bonds between the coating and the underlying metal may be thought of as elastic springs which fail irreversibly under excessive tensile stress (see figure 2(d)).

This suggests the presence of a parameter which describes the failure stress $\sigma_{ad}^*$ and a debonding rate which is driven by the tensile stress $\sigma_{ad}$. Finally, we speculate, based on considerations of simplicity, that this rate is linear with respect to the driving force beyond $\sigma_{ad}^*$.

Anticipating potential confusion in word use, we note that the Heaviside function $H(\cdot)$ renders equation 6 nonlinear, even if the equation is linear for both $\sigma_{ad} < \sigma_{ad}^*$ and $\sigma_{ad} > \sigma_{ad}^*$. We thus refer to the growth rate as irreversible and nonlinear.

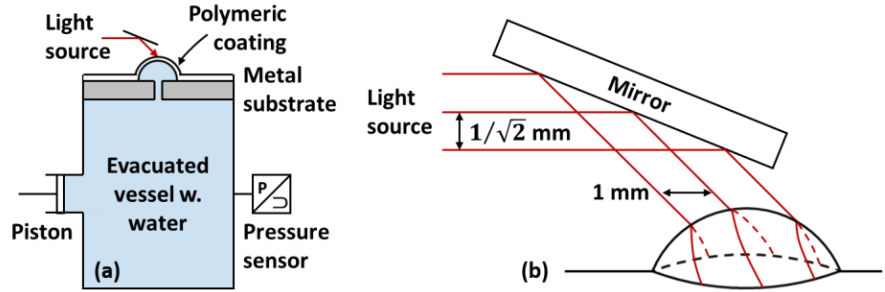

Figure 3: A simplified schematic of the experimental setup used by [35]. (a) The setup consists of an airtight deaerated vessel containing water. A piston, used to modify the pressure within the vessel, and a pressure transducer, used to track the pressure within the vessel as a function of time, are attached to the vessel. The top of the vessel is clamped shut with a coated metal substrate. The metal substrate possesses a pinhole defect of known size. Upon application of pressure, the blister deforms and detaches from the underlying substrate. (b) The curvature of the blister is tracked using an optical device consisting of parallel planar light sources of known separation. The pattern formed on the surface of the blister is recorded and processed assuming that the blister is a sphere segment. Van der Meer-Lerk and Heertjes recorded the average radius, the pressure, and the contact angle as a function of time. For a more complete discussion on the experimental setup, which includes vessel evacuation and substrate replacements / reuse, see [35].

The data validating equation 6 is taken from [35]. A simplified schematic of the experimental setup is shown in figure 3. In brief, the setup consists of an airtight deaerated vessel containing water. The top of the vessel is clamped shut with a coated metal substrate; the metal substrate possesses a pinhole defect of known size. By displacing a piston attached to the vessel, the pressure in the vessel is temporarily raised, causing the coating to deform and detach from the underlying substrate about the pinhole defect. Coating deformation and blister growth causes the pressure to relax, and the pressure of the vessel is tracked as a function of time using a pressure transducer. An optical device consisting of parallel planar light sources of known separation is used to track the height and the contact angle as a function of time.

The experiment was performed using stainless steel substrates. The coatings investigated were polyurethane, epoxy, alkyd, and chlorinated rubber, although van der Meer-Lerk and Heertjes (the authors of [35]) were only able to obtain meaningful quantitative polyurethane and alkyd data, the remainder breaking before a sufficiently large data set could be gathered. The stated intention of their work was to investigate a parameter describing the adhesion strength of coating / substrate pairs. In this regard, our aims overlap with theirs; however, our definition of critical adhesion stress $\sigma_{\mathrm{ad}}^{*}$ is more comprehensive, as it accounts for the effect of deformation at the perimeter of the blister. Furthermore, our theoretical approach differs considerably from that of van der Meer-Lerk and Heertjes, who employed a graphical method reliant on visual extrapolation to zero rate of change of blister area.

In the present work, we have re-processed the polyurethane and alkyd data using equations 11 and 15. Combining these two equations together, we get:

$$\sigma_{\mathrm{ad}} = \frac{\Pi r}{2l(t=0)} \frac{2(1-\cos\psi)}{\sin^2\psi}$$

(27)

We then plot $dr/dt$ as a function $\sigma_{\mathrm{ad}}$ to determine $\sigma_{\mathrm{ad}}^{*}$ and validate the form of equation 6.

### 3.2. Deformation Under Osmotic Pressure

Our second validation data set is taken from an earlier work by the same pair of authors [25]. In brief, droplets containing known amounts of magnesium acetate, potassium hydrogen sulfate, zinc sulfate, or sodium sulfite were deposited on top of polished, degreased stainless steel surfaces. These salts leave behind uniform deposits, which were subsequently coated over with polyurethane, epoxy, or chlorinated rubber. The coated substrates were then immersed in distilled water for 160 days. At irregular intervals, the coated substrates were removed from water, wiped dry, and analyzed using a light-section microscope. The maximum height and the size of the major and minor axes of the blisters were recorded. A summary of the experimental procedure is shown in figure 4.

This data set differs from their later work in several respects. Osmotic pressure, effected through entrapped soluble salts, was used in place hydrostatic pressure, which provides a more direct validation of the present work. The accuracy of the measurement was also improved through the use of light-section microscopy, and the analysis allows for ellipsoidal segments, which retains more information concerning the surface area and volume of the blister.

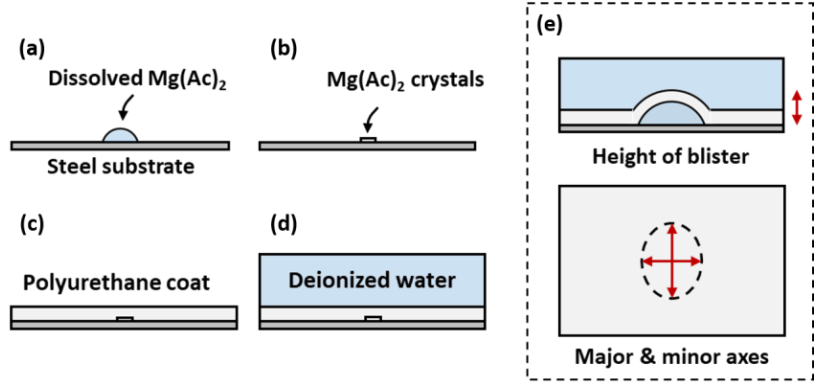

Figure 4: A summary of the experimental procedure in [25]. In chronological order, (a) a known amount of salt dissolved in water is deposited on top of a polished, degreased stainless steel substrate, (b) the solution is allowed to evaporate, leaving behind a uniform layer of salt, (c) a coating is cured on top of the salt, and (d) the coated substrate is immersed in distilled water. (e) The height and the size of the major and minor axes are recorded at irregular intervals using light-sectioning microscopy. Note that the authors have only reported quantitative blistering data sets for magnesium acetate / polyurethane pairs; this has been reflected in the figure.

However, the only blistering data sets which were reported quantitatively by van der Meer-Lerk and Heertjes (in figure 2 of [25]) are the blister volume versus time data for three different amounts of magnesium acetate. Our attempts to recover the remaining data sets are not successful, as the data from the laboratory is lost. We extract the blister volume versus time data from figure 2 of [25] using the *grabit* toolbox [50], downloaded from MATLAB central. This data set has been included in the supplementary information for posterity; we recommend not repeating the data extraction procedure using the figures presented in our work, to avoid error accumulation.

The experimental procedures reported in figure 4 requires some modification on the basic model reported in table 2 and some deductive work based on other data sets reported by van der Meer-Lerk and Heertjes. The model developed in section 2 assumes that the transport of water is fast relative to the soluble species. This is needed to obtain a well-defined expression for osmotic pressure, which allows the curvature of the blister to be directly evaluated from equation 11. However, the data sets presented by van der Meer-Lerk and Heertjes were obtained at timescales in which the transport of water dominates coating deformation, and osmotic pressure is not well-defined. We thus modify the basic model by eliminating equations 16 and 11. The former is replaced by a mass balance for water, i.e., equation 17 and the latter is replaced with a thermodynamic expression:

$$V = V(n_i, T, P) \approx \sum_i n_i \tilde{V}_i \qquad (28)$$

Here $\bar{V}_i$ is the molecular volume of pure species $i$. More exact expressions for $V(n_i, T, P)$ can be used, but this suffices for the present work, as most blister fluids of practical interest are dilute.

The diffusivity of water in the polyurethane coating, the Young's modulus of the coating, and the critical adhesion stress of the coating / substrate pair are the missing parameter values. These parameter values are expected to influence the accuracy of the model substantially, but are strong functions of the coating composition and surface preparation method, which are not comprehensively discussed in the authors' body of work. Since the authors are no longer contactable, we did the next best thing, which is to infer parameters values from the authors' other works. The diffusivity of water is obtained by fitting the steady-state permeation data reported in figure 5 of [25] to the following expression:

$$\frac{1}{D_{H_2O}} = k_1(1 - x_{H_2O}) + k_2 \tag{29}$$

Here $k_1$ and $k_2$ are empirical fitting parameters, and $x_{H_2O}$ is the mole fraction of water. This data set is obtained though Payne cup measurement and is independent of the validation data set. Refer to the aforementioned work for experimental details on the Payne cup measurement.

Equation 29 is inspired by the Stokes-Einstein relation, wherein the diffusivity of the dilute soluble species is inversely proportional to size:

$$\frac{1}{D_{H_2O}} \sim r_{H_2O} \tag{30}$$

Here $r_{H_2O}$ is the size of water molecules. Water molecules tend to cluster about the soluble species, which increases the effective size of the species diffusing through the polymer matrix. The simplest form consistent with this expression is:

$$\frac{1}{D_{H_2O}} \sim 1 - x_{H_2O} \tag{31}$$

Finally, we acknowledge that the Stokes-Einstein relation only holds for dilute soluble species, whereas water is the dominant species of the blister fluid. In particular, it must have a finite self-diffusion coefficient within the coating, giving us the form of equation 29. The fitting result for the diffusivity of water in polyurethane is shown in figure 5(a). The data used to fit the diffusivity of water in polyurethane is recorded for posterity in the supplementary information.

The Young's modulus is obtained by fitting data from van der Meer-Lerk and Heertjes' later work [35]. In brief, we solve equation 11 for each data point reported in table 1 of [35] assuming $K_{MR} = -s_1/s_{-1} = 1.5$ [54]. The Young's modulus is obtained from the limiting expression:

$$E = 6s_1(1 + K_{MR}) \qquad (32)$$

The Young's modulus is then plotted against the strain rate:

$$\dot{\gamma} = \frac{1}{r}\frac{dr}{dt} \qquad (33)$$

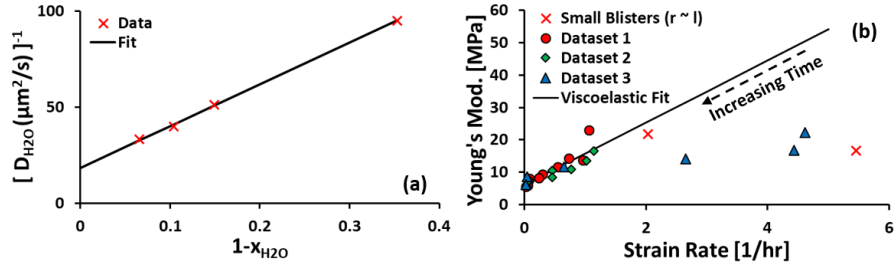

Figure 5: (a) The semi-empirical fit for the diffusivity of water in polyurethane. The equation of the semi-empirical fit is $D_{H_2O}^{-1} \, [\mathrm{s}/\mu\mathrm{m}^2] = 216.7(1 - x_{H_2O}) + 18.4$. The trend of increasing mobility with decreasing fraction of soluble components is consistent with observations made in the literature [51, 52, 53] for a fairly broad range of solute-solvent pairs. (b) The Young's modulus of polyurethane is found to vary linearly with strain rate at low strain rates. In the present work, and in most cases of practical interest, the strain rate is low, which gives the limiting value $E = 6.2$ MPa.

The result is shown in figure 5(b). Linear regression is performed for strain rates $\leq 1.5 \; \mathrm{hr}^{-1}$; the dependence of Young's modulus on strain rate is a clear sign of viscoelastic behavior. For the timescale of practical interest ($\sim 24 \; \mathrm{hr}$), the strain rate is expected to be negligible, which validates the elastic assumption made in section 2. We note that the data is heteroscedastic about the best-fit line, which arises because the strain rate is inhomogeneous along the surface of the coating. The smallest strain rates correspond to coatings which have relaxed for the longest duration, for which the heterogeneity is expected to have vanished.

The critical adhesion stress is obtained from the first validation data set discussed in section 3.1. The set of parameter values used for comparison against the second validation data set is listed in table 1.

### 3.3. Critical Delamination Length

Our third validation data set, generated specifically for the present work, is obtained in relation to three intertwined hypotheses. Combining equation 14 and equation 27, and assuming that adhesion stress exceeds critical adhesion stress, we obtain:

$$\frac{dr}{dt} = k_{ad}\left(\frac{\Pi r}{2l(t=0)}\frac{2(1-\cos\psi)}{\sin^2\psi} - \sigma_{ad}^*\right) \qquad (34)$$

Ordinarily, the osmotic pressure $\Pi$ is inversely proportional to volume ($\sim r^3$) and thus the blister growth stops. However, if the corroding surface can act as a source of soluble materials, then the osmotic pressure may not decrease with $r^3$. Two possibilities arise here - if the solubility limit is not exceeded, then new soluble materials are produced at a rate proportional to the total surface area of the blister ($\sim r^2$). Then at $t \to -\infty$,

$$r \sim \sqrt{t} \tag{35}$$

| Symbol | Value | Unit | Note |
|---|---|---|---|
| $C_{i,\text{out}}$ | 0 | mol·m$^{-3}$ | |
| $C_{\text{H}_2\text{O},\text{out}}$ | 55600 | mol·m$^{-3}$ | |
| $D_{\text{H}_2\text{O}}^{-1}$ | $216.7(1 - x_{\text{H}_2\text{O}}) + 18.4$ | s/$\mu$m$^2$ | See figure 5(a) |
| $E$ | $6.2 \times 10^6$ | Pa | See figure 5(b) |
| $k_{\text{ad}}$ | $2.07 \times 10^{-12}$ | m/s/Pa | See figure 8(a) |
| $K_{\text{MR}}$ | 1.5 | | See [54] |
| $l(t = 0)$ | 80 | $\mu$m | See [25] |
| | $8.81 \times 10^{-9}$ | | Case 1, [25] |
| $n_{MgAc_2}(t = 0)$ | $2.98 \times 10^{-8}$ | mol | Case 2, [25] |
| | $3.96 \times 10^{-8}$ | | Case 3, [25] |
| | 107 | | Case 1, [25] |
| $r(t = 0)$ | 161 | $\mu$m | Case 2, [25] |
| | 177 | | Case 3, [25] |
| $t_{\text{span}}$ | 160 | days | See [25] |
| $\sigma_{\text{ad}}^*$ | $2.5 \times 10^6$ | Pa | See figure 8(a) |
| $T$ | 298 | K | |

Table 1: Table of parameter values used for comparison against the second validation data set. The initial blister size $r(t = 0)$ is calculated based on the volume of magnesium acetate droplet at the saturation limit (see [55]), assuming a 70° contact angle (see [56]). All parameter values are obtained independently of the second validation data set.

Otherwise, the osmotic pressure is set by the solubility limit and is thus approximately constant, with an exponential limiting growth rate:

$$r \sim \exp(t) \tag{36}$$

We note that this analysis ignores the effect of hydrostatic pressure, which scales with $r$, but only if the blister is not immersed in electrolyte, which is beyond the scope of the present work.

Assuming that osmotic pressure is set by the solubility limit, a critical length scale emerges:

$$r_{\text{de}}^* = \frac{2l(t=0)\sigma_{\text{ad}}^*}{\Pi_{\text{sat}}} \frac{\sin^2 \psi}{2[1-\cos\psi]} \sim \frac{l(t=0)\sigma_{\text{ad}}^*}{k_B T C_{\text{ref}}} \tag{37}$$

We call this length scale the critical delamination length, defined as the blister size above which unstable growth occurs, leading to delamination. We hypothesize that coating delamination in field applications arise due to this inherent instability.

It is immediately evident to us that $Fe(OH)_3$ cannot give rise to plausible critical delamination length values. The solubility product of $Fe(OH)_3$ ($\sim 10^{-38}$ $M^4$ [57]) gives rise to mPa-range osmotic pressure and km-range critical delamination lengths. More probable results are obtained with $Fe(OH)_2$, whose solubility product ($\sim 10^{-15}$ $M^3$ [57]) leads to Pa-range osmotic pressure and m-range critical delamination lengths.

This is still large but is calculated assuming a critical adhesion stress of $\sim 1$ MPa. There is a consensus within the organic coating literature that the critical adhesion stress of broad classes of polymeric coatings is lowered when corrosion occurs [23, 24, 21, 58]. This mechanism is termed coating saponification and occurs when ester bonds within the coating break in the high-pH environment of the blister:

$$R_1COOR_2 + OH^- \rightarrow R_1COO^- + R_2OH \tag{38}$$

Here $R_1$ and $R_2$ are hydrocarbon groups. For alkyd and acrylic coatings, ester bonds are located within the backbone and the side chain of the polymer, respectively. For epoxy coatings, ester bonds may be present depending on the choice of curing agent.

We thus have three hypotheses. First, corrosion-induced osmotic pressure can cause unstable delamination for blisters above a certain size. Second, $Fe(OH)_3$ cannot be reasonably expected to cause unstable delamination in field applications - however, $Fe(OH)_2$ can. Third, this event is made more probable if there is significant loss of adhesion due to saponification. Ultimately, we were only able to confirm the second hypothesis. However, it seems valuable to discuss our setup, if only to stimulate discussion of potential working variations. We wish to create blisters of known size; from there, we can study the rate of blister growth as a function of its initial size. Our setup consists of 4 by 6 in. type-R Q-panels [59]. On each panel, we apply a circular patch of DuPont nonstick dry film lubricant with Teflon [60]. The circular patch is applied using a small ($\sim 1$ mm) cotton swab which has been immersed for 24 hours in excess DI water to remove water-soluble components and dried at room temperature and pressure for a further 24 hours. The lubricant evaporates almost immediately upon application, creating a thin but uneven layer of Teflon which is smooth to the touch. The average thickness of the Teflon layer is calculated from its mass, density, and surface area, and is found to be consistently in between 2 and 4 $\mu m$ across all panels. At this thickness, the Teflon layer is not expected to provide significant mass-transfer resistance relative to the coating.

Afterwards, 200 $\mu$m (wet thickness) water-based coatings are applied using a BYK-Gardner steel drawdown bar. Two commercial acrylic latex anti-corrosion coating formulations (pigmented and non-pigmented), generously provided by Dow Chemical Company, are considered in the present work. The coatings are allowed to dry for 24 hours at room temperature and pressure. After the coatings dry, we affix a 300-mL VWR funnel glass centered at the circular patch of Teflon using Gorilla clear silicone sealant. The sealant is applied over the outer edge of the funnel glass and allowed to dry at room temperature under pressure ($\sim$500 Pa above atmosphere). The funnel is then filled with 100 mL of DI water equilibrated with air. The top view of the blister is photographed using a Samsung Galaxy S5 at 3, 6, 12, 18, 24, and every 12 hours hence up to 10 days.

Every 24 hours, 25 mL of DI water is siphoned off from each container, and the solution is topped up to 100 mL with equilibrated DI water to ameliorate oxygen and water depletion due to corrosion and evaporation, respectively. A schematic of the setup is shown in figure 6.

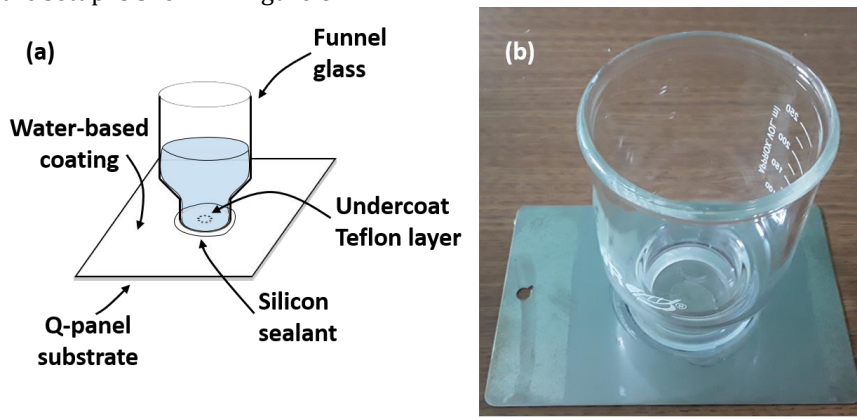

Figure 6: (a) Schematic of the experimental setup and (b) a physical realization of the setup.

This setup is exploratory in nature, in line with our goal of creating blisters with known initial size. We also explore the effect of adding a pinhole defect at the center of the coating using a 25G medical syringe as a way to initiate blister formation [23]. In all, a total of four factors are explored, namely (1) pigment versus no pigment, (2) Teflon versus no Teflon (i.e., control), (3) diameter of circular Teflon patch ($\sim$0.5 cm versus $\sim$1.5 cm), and (4) defect versus no defect. For each experiment, we perform a replicate, resulting in a total of 24 panels. The list of experiments and a selection of experimental results are shown in the supplementary information.

## 4. Model Predictions and Validation Results

Our model predicts the qualitative behavior summarized in figure 7. At time t = 0, blisters initiate at sites exhibiting a combination of low critical adhesion stress $\sigma_{\text{ad}^*}$ and high solute concentration. Localized loss of $\sigma_{\text{ad}^*}$ may occur due to interface contamination by oils, salts, solids, or other inclusions [6] which prevent the coating from adhering to the surface. In particular, we note that uneven surfaces, in conjunction with low metal / coating surface energy, encourages entrapment of air bubbles, which behave like unprotected cavities. More generally, coating formulations with low metal / coating surface energy are expected to have low $\sigma_{\text{ad}^*}$ and blister with relative ease. Interface contamination by salts also introduces soluble components which can generate osmotic pressure which drives osmotic blistering. These explain the field observation that coatings tend to last longer when applied on polished, degreased surfaces.

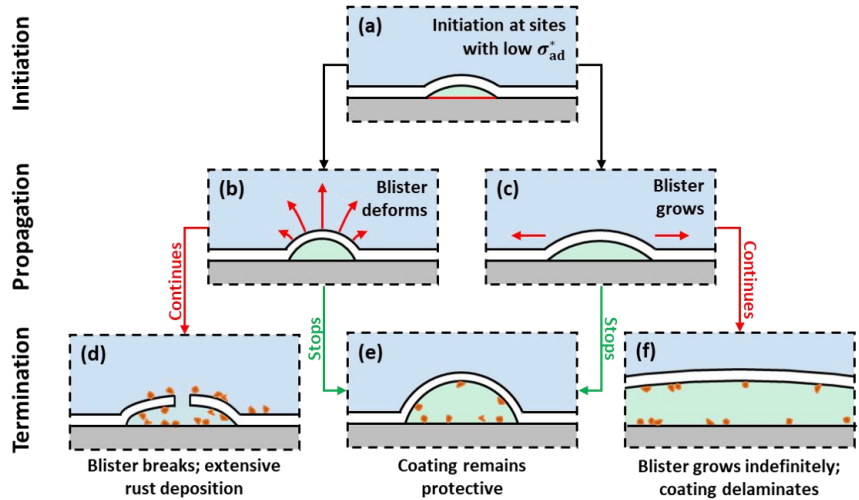

Figure 7: (a) Blisters initiate at sites with low critical adhesion stress and / or high solute concentration. (b) Blisters propagate by deforming, thus increasing its contact angle, and / or (c) by growing, thus increasing its radius. (d) Deformation may cause the coating to exceed its maximum elongation, which then breaks. If this happens, the exposed steel surface corrodes freely, resulting in extensive rust formation. (e) In the ideal case, deformation and growth stops after some time, and the coating remains protective, i.e., acts as a barrier to transport. Sufficiently small blisters remain invisible to the naked eye, which preserves the aesthetics of the coating. (f) Sufficiently large blisters grow in an unstable manner, causing eventual delamination.

Blisters can propagate by deformation and / or growth, as shown in figure 7(b) and (c), respectively. Deformation occurs when adhesion is sufficiently strong to hold the perimeter in place - transport of water causes the blister to swell instead, and the contact angle increases. Growth occurs when adhesion stress exceeds its critical value, and the base radius increases. If deformation and growth stop, the coating remains protective (figure 7(e)); as discussed in section 5, oxidation of Fe$^{2+}$ to Fe$^{3+}$ occurs and a protective oxide layer builds up. This reduces the rate of corrosion, and shifts the solution equilibrium to that of Fe$_2$O$_3$, which does not generate enough osmotic pressure to deform the blister. The blister then deflates. If deformation continues unabated, the maximum elongation of the coating is eventually exceeded (see equation 5), and the blister breaks (figure 7(d)). If growth continues unabated, the critical delamination length is eventually exceeded, and the coating grows in an unstable manner (figure 7(f)), assuming that the surface does not passivate.

### 4.1. Endpoint Classification

The endpoints summarized in the last row of figure 23 can be more comprehensively classified by asking three questions:
1. Does deformation stop?
2. Does growth stop?
3. What is the source of osmotic pressure?

In what follows, we discuss each of these possible outcomes and the dimensionless numbers that govern them. The limitations of this classification are discussed in sections 5.1 and 5.2.

#### 4.1.1. Case 1: deformation does not stop

If the osmotic pressure arises from entrapped soluble components, then blister growth must stop at some point, due to the scaling arguments presented in section 3.3. Assuming that the coating is immersed in water, we obtain a scale for the base radius and the osmotic pressure:

$$\overline{r}(\overline{t} \to \infty) \sim \sqrt{\frac{\sum_i \overline{n}_i}{\overline{\sigma}^*_{\mathrm{ad}}}} \tag{39}$$

$$\overline{\Pi}(\overline{t} \to \infty) \sim \sqrt{\frac{\left(\overline{\sigma}^*_{\mathrm{ad}}\right)^3}{\sum_i \overline{n}_i}} \tag{40}$$

Using these, and the third and fourth equations in table 2, we estimate the contact angle $\psi$.

If the osmotic pressure arises from corrosion products, and assuming that the solubility limit of the corrosion products is exceeded, then the osmotic pressure is constant. This osmotic pressure is unlikely to cause critical delamination, due to the small initial size of the blister. We thus expect:

$$\overline{r}(\overline{t} \to \infty) \sim \overline{r}(\overline{t} = 0) \tag{41}$$

$$\overline{\Pi}(\overline{t} \to \infty) \sim \frac{\sum_i \overline{n}_i}{[\overline{r}(\overline{t}=0)]^3} \tag{42}$$

As before, using these, as well as the third and fourth equations in table 2, we estimate $\psi$.

With $\psi$ available, we calculate the dimensionless number:

$$\pi_{\text{break}} = \frac{1}{\epsilon}\tan^2\left(\frac{\psi}{2}\right) \tag{43}$$

This dimensionless number describes the tendency for a blister to break. In particular,

$$\pi_{\text{break}} = \begin{cases} \gg 1 \rightarrow \text{Blister breaks} \\ \ll 1 \rightarrow Blister\ does\ not\ break \end{cases} \tag{44}$$

Note that the first line of equation 44 is a trigonometric simplification of the fifth equation in table 2.

### 4.1.2. Case 2: growth does not stop

As discussed prior, if osmotic pressure is only generated by soluble components, then blister growth must stop at some point. Consequently, the coating cannot be said to be delaminated, i.e., completely detached from the underlying substrate, aside from the subjective aesthetic element of a loosely attached coating.

Let $\bar{r}_{\text{aesthetic}}$ be the size of a blister which is visibly displeasing. Then, if osmotic pressure is generated by soluble components, we can evaluate the dimensionless number:

$$\pi_{\text{aesthetic}} \equiv \bar{r}_{\text{aesthetic}}\sqrt{\frac{\bar{\sigma}_{\text{ad}}^*}{\sum_i \bar{n}_i}} \tag{45}$$

This dimensionless number describes the tendency for the blister to be visibly displeasing:

$$\pi_{\text{aesthetic}} = \begin{cases} \gg 1 \rightarrow \text{Blister is displeasing} \\ \ll 1 \rightarrow \text{Blister is not displeasing} \end{cases} \tag{46}$$

On the other hand, if osmotic pressure is generated by corrosion products, then we encounter the case of critical delamination length, as described in section 3.3 and in equation 37. The following dimensionless number arises:

$$\pi_{\text{delamination}} \equiv \frac{\bar{r}(\bar{t}=0)}{\bar{r}_{\text{de}}^*} \tag{47}$$

This dimensionless number describes the tendency of the coating to undergo critical delamination:

$$\pi_{\text{delamination}} = \begin{cases} \gg 1 \rightarrow \text{Blister undergoes unstable growth} \\ \ll 1 \rightarrow \text{Blister does not grow} \end{cases} \tag{48}$$

### 4.1.3. Case 3: deformation and growth do stop

This case is trivial, as it is characterized by the negation of cases 1 and 2, i.e., $\pi_{\text{break}} \ll 1$, $\pi_{\text{aesthetic}} \ll 1$, and $\pi_{\text{delamination}} \ll 1$.

### 4.2. Model Validation

The validation results corresponding to section 3.1 and section 3.2 are summarized in figure 8(a, b) and (c, d), respectively. The fit in figure 8(a) is excellent, but there is a tendency to underestimate growth rate at low adhesion stress. We considered the possibility of a quadratic dependence on stress difference:

$$\frac{dr}{dt} = k_{\text{ad}}(\sigma_{\text{ad}} - \sigma_{\text{ad}}^*)^2 H(\sigma_{\text{ad}} - \sigma_{\text{ad}}^*) \tag{49}$$

This quadratic dependence could arise due to viscoelastic delay, which can be significant within the timescale of the experiment. To test this hypothesis, we performed a $\chi^2$ test using equation 15 as the null hypothesis. This leads to a $p$-value of 0.23 and rejection of the quadratic dependence.

Equally good fits are obtained using the alkyd data set (figure 8(b)), but only if the data is split into the first pressure step and all subsequent pressure steps. We speculate that there is a considerable time gap between these two subsets of data, during which the interface changes considerably by wetting or corrosion, although this is difficult to verify, as the authors are no longer contactable.

The model prediction in figure 8(c) shows good agreement with the data, especially considering that the model prediction is made completely a priori, i.e., with no fitting parameters. For $t < 50$ days, the model under-predicts blister volume and, otherwise, over-predicts it. We suspect that this is caused by an incorrect diffusion model. As shown in figure 8(e), the data set shows a linear dependence on $\sqrt{t}$ for most of the duration of the experiment. This suggests that the diffusivity is constant, in contrast to the result obtained from the Payne cup experiment. A greatly improved fit is observed when $D_{\text{H2O}}$ is used as a fitting parameter, as seen in figure 8(d).

The fitted value of $D_{\text{H2O}}$ is found to be $3.36 \times 10^{-14}$ m$^2$/s, which corresponds to $x_{\text{H2O}} \approx 0.95$ using the correlation in figure 5(a). We speculate that the failure of the correlation is caused by the unsteady-state nature of the experimental setup in figure 4. At time t = 0, the coating immersed in DI water is dry; the salt is trapped underneath the coating, and there should be no salt within the coating. Water thus penetrates from the outside into the inside of the blister in its pure state. When the water front begins wetting the salt, the salt dissolves and goes into the coating. This logic suggests that, for most of the early part of the experiment, the average concentration of salt within the pore volume of the coating would be lower than the idealized value of $n_i/2V$ present in the steady-state Payne cup experiment. We believe that the correlation in figure 5(a) is valid, but the proper value of $x_{\text{H2O}}$ requires modelling of unsteady-state transport through a polymer matrix, which is beyond the scope of the present work.

As discussed in section 3.3, we were ultimately unable to experimentally verify the existence of a critical delamination length. However, as shown in figure 9, there is a clear dark-green patina characteristic of mixed $Fe^{2+}$ / $Fe^{3+}$ ions, with a tendency towards lighter green coloration near the delaminating edge of the blister indicating a higher proportion of $Fe^{2+}$. This is in general agreement with the qualitative prediction of the model summarized in figure 7. As the blister progresses, corrosion produces $Fe(OH)_2$, which appears as a green patina. A combination of osmotic pressure arising from corrosion products and saponification due to increased pH causes the coating to detach from the underlying metal substrate. The only aspect which has not been accounted for in the model is the tendency for $Fe^{2+}$ to oxidize to $Fe^{3+}$ near the center of the blister. We speculate on two possible explanations - rust precipitation at the center of the blister could create a protective oxide layer which reduces corrosion, thus increasing the availability of oxygen for oxidizing $Fe^{2+}$. Alternatively, this could be a consequence of chemical kinetics, i.e., it takes time for $Fe^{2+}$ to be oxidized to $Fe^{3+}$. In either hypothesis, it makes sense for a metallic surface freshly exposed by the delamination process to show a lighter green coloration.

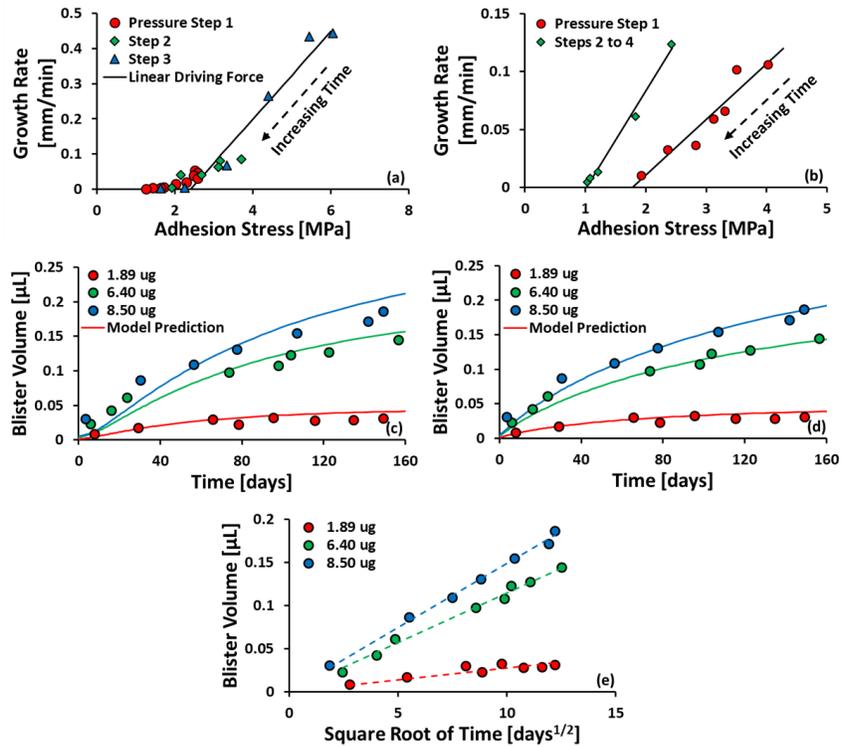

Figure 8: (a,b) Plot of growth rate as a function of adhesion stress for steel / polyurethane and steel / alkyd, respectively. Adhesion stress is calculated using equation 15 from the raw data published in [35]. Solid black lines are the fit from equation 14. (c) Plot of blister volume as a function of time. Model prediction has no free parameters, i.e., no fitting is performed. Data is taken from [25]. (d) Same as the previous plot, but the diffusivity of water in the coating is assumed to be constant w.r.t. composition and fitted to the data set. The resulting diffusivity is $3.36 \times 10^{-14}\,\mathrm{m^2}/s$, equivalent to the diffusivity at $x_{H2O} \approx 0.95$ based on the correlation in figure 5(a). (e) Plot of blister volume as a function of $\sqrt{t}$. The dotted lines are best-fit linear equations passing through (0,0).

At first glance, the progression shown in figure 9 seems to support the critical delamination hypothesis, in particular the scaling relation $r \sim \exp(t)$, as the radius of the undercoat rust pattern increases more rapidly with time. However, as discussed in the supplementary information, the Teflon layer seems to impart a significant degree of corrosion resistance. As a consequence, it is not obvious whether the accelerating growth is due to increasing osmotic force, which scales with the surface area or due to the corroding area spreading beyond the Teflon patch.

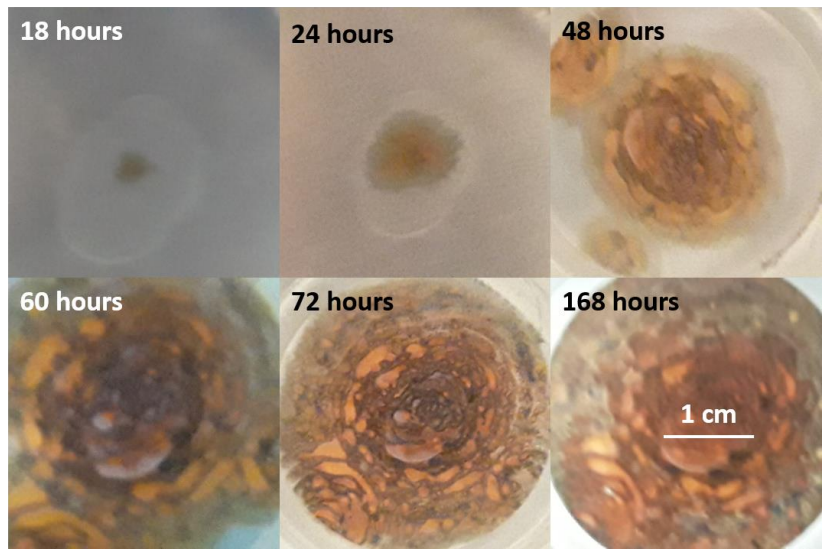

Figure 9: Development of undercoat blister and rust. This set of figures correspond to experiment 9 as listed in the supplementary information. Dark-green coloration is observed at the edge of the blister at all points in time, at least until the blister edge meets the funnel glass. As the blister grows, small pocket-shaped patterns develop. By hour 168, the surface coloration has turned mostly brown. Similar observations were made for other experiments listed in the supplementary information; for brevity, these images are placed in the supplementary information.

The formation of pocket-shaped patterns radiating outward from the initial defect suggests that coating delamination occurs in batch. In context of the present work, as osmotic pressure builds up, the coating deforms, causing the contact angle to be more obtuse. Following equation 27, the adhesion stress increases until it exceeds the local critical adhesion stress. If the critical adhesion stress is inhomogeneous, then the boundary slips and the contact angle drops. The process then repeats itself, resulting in the pattern observed in figure 9. If this description is accurate, we can further deduce that the length scale of critical adhesion stress inhomogeneity for our metal / coating pair is $\sim 100\ \mu$m.

## 5. Discussion

In this section, we discuss the limitations and applications of the model within and beyond the field of corrosion.

### 5.1. Effect of Surface Passivation on Model Prediction

As illustrated in figure 9, a considerable amount of $Fe_2O_3$ is precipitated at the center of the blister. To what degree does this rust layer protect against corrosion and how does that impact the model predictions?

To our knowledge, even to present day, there is still uncertainty on the degree to which rust precipitation protects against corrosion. On the materials engineering side, the well-known Pilling-Bedworth theory [61] predicts a porous unprotective oxide layer, if the oxide is $Fe_2O_3$, and a dense protective oxide layer, if the oxide is $Fe_3O_4$, provided that corrosion reaction is sufficiently rapid. Current studies suggest that, aside from the edge of the blister, the oxide produced by undercoat corrosion of iron contains mostly $Fe_2O_3$ [23], a conclusion which is supported by the undercoat visualization results shown in figure 9 and more clearly in the photo-micrograph taken by Hare [24]. However, other studies have suggested that underneath this layer of $Fe_2O_3$, there is a thin layer of $Fe_3O_4$ [62].

Furthermore, due to finite oxidation and precipitation kinetics [48], one would expect rust to precipitate away from the surface of the metal. This follows from the following sequence of events: $Fe^{2+}$ ions are produced by corrosion, oxidizes in the bulk volume of the blister to $Fe^{3+}$, and then precipitates on the exposed surface. One would then expect the rust layer to be porous regardless of the Pilling-Bedworth ratio, a conclusion which is supported by high-temperature iron corrosion experiments [63]. However, the very same set of experiments also show the formation of a thin, dense, protective oxide layer underneath a loose, porous one. This is attributed to the Arrhenius dependence of corrosion rate on temperature, which causes the formation of a large excess of $Fe^{2+}$ and $Fe^{3+}$ ions which then precipitate within the pore volume of the porous oxide layer.

Electrochemical experiments support the result of the aforementioned high temperature iron corrosion experiment. The existence of a passivation potential / regime for corroding steel is well-known; in brief, as surface potential is increased, instead of an exponentially increasing corrosion rate, as expected from Butler-Volmer theory, the corrosion rate nearly vanishes when one goes above the passivation potential [45, 64]. This suggests that very fast corrosion produces excess $Fe^{2+}$ and $Fe^{3+}$ ions which precipitate as a dense, protective layer. However, this passivation regime can quickly vanish when a small amount of $Br^-$, $Cl^-$, $SO_4^{2-}$, etc. (i.e., anions of strong acids) is added to the electrolyte [65, 62, 66].

Of course, the identity of the metal matters as well. Extremely reactive metals can be very passive, as is the case of aluminum. For steel, there is a fairly broad range of reactivity, depending largely on the composition of the alloy. Stainless steel would be expected to give low osmotic pressure but no passivation; magnesium-iron alloy would be expected to give high osmotic pressure but significant passivation. In the middle, an unprotective porous oxide layer is expected, although we note that even a porous oxide layer can impart some degree of protection by surface occlusion [22].

### 5.2. Coating Defects

The model we present assumes a smooth, isotropic coating. In practice, coatings almost always exhibit some form of defect, introduced during application or use, e.g., air bubbles during spraying, rugosity due to the bristles of the brush used, and scratches from daily use. These impact the blistering process by providing small open sites where corrosion can take place - the resulting saponification reactions, if applicable, then causes localized loss of $\sigma_{ad^*}$ , thus initiating osmotic blistering. This phenomenon is clearly observed in figure 10, wherein a scratch deliberately introduced to a coating causes the formation of blisters near the scratch, possibly compounded by loss of adhesion due to mechanical damage. At the same time, the presence of such defects can act as locations where ions leak, removing the osmotic pressure needed to cause blistering, provided that the defects are not sealed by the corrosion products [23].

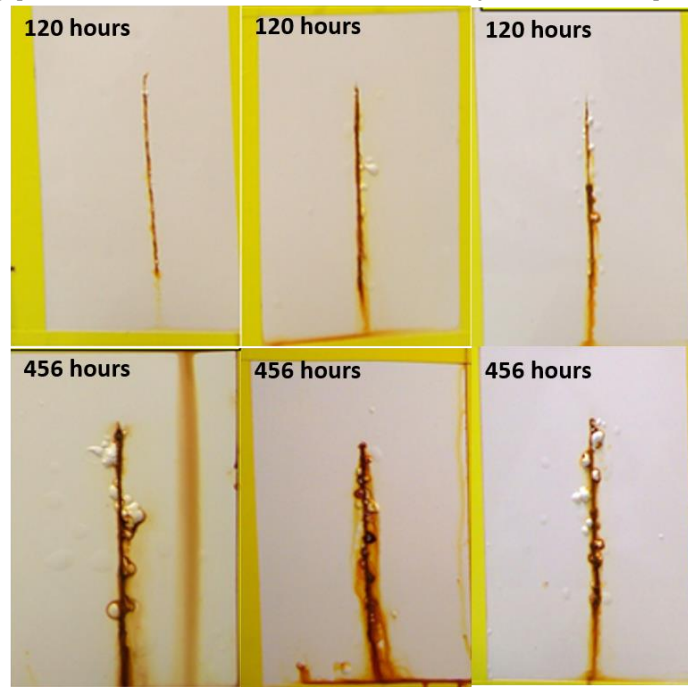

Figure 10: Blisters develop about a scribe introduced to three different coating formulations undergoing salt spray test. This progression is in agreement with the qualitative description of a surface undergoing saponification about the scribe, followed by blistering driven by osmotic pressure generated by corrosion products. However, for this to happen, defects must become sealed by corrosion products.

In short, there is considerable uncertainty regarding the nature of surface passivation and coating defects. Given that most practical setups incorporate some amount of aggressive anions and operate at corrosion potential, there should be a lack of passivation in most field applications. If the surface does passivate, or if there is a large pinhole defect which allows the content of the blister to escape, then critical delamination does not occur; otherwise, the qualitative conclusions of the study is unchanged. However, as far as conservative design is concerned, critical delamination should be considered.

*5.3. Filiform Corrosion*

The present work has been largely focused on osmotic blistering, but similar explanations can also be used to explain other corrosion phenomena. Filiform corrosion refers to the formation of finger-like patterns on coated aluminum [23] starting at some macroscopic defect sealed by semi-permeable corrosion products. The incumbent explanation assumes that the coupling between corrosion, saponification, and transport phenomena results in unstable growth at the tips of the finger. As corrosion progresses, $OH^-$ ions are produced, which causes saponification (see reaction 38). The $OH^-$ ions are lost into the surrounding electrolyte by diffusion, in particular near the semi-permeable defect, which acts as a hole through which ions can escape with relative ease. This results in an increased concentration of $OH^-$ ions far away from the defect, in particular at the tips of the finger. The tips thus preferentially debond, causing the filiform pattern.

Other hypotheses, e.g., anodic undermining due to formation of differential aeration cell [67], electro-osmotic forces acting normal to the metal /coating interface [68], inertial forces due to fluid flow driven by osmosis [69] have also been put forward as possible explanations for filiform corrosion. We note that the saponification, anodic undermining, and electro-osmotic hypotheses do not provide a mechanism explaining the direction of filiform growth; only that filiform growth is enhanced at the tip of the fingers. In context of the present work, this direction can be determined by inhomogeneity in critical adhesion stress. If this is the case, then we expect the length scale of the inhomogeneity to be the width of the fingers. For the inertial force hypothesis, inhomogeneity may play a considerable role in causing sudden changes in the direction of filiform growth, in which case we expect the length scale of the inhomogeneity to be the distance between sudden changes in growth direction.

The model presented in section 2.1 is appropriate for studying filiform corrosion, provided that some spatial refinement is included, e.g., by breaking the fingers into compartments with its own mass balance. Due to the similarity in the underlying physics, we expect the conclusions drawn in section 4.1 to be transferable to filiform corrosion, which allows prediction of unstable growth, coating rupture, and filiform visibility.

### 5.4. Other Applications of the Model

The particular use of metal / coating pairs is not intrinsic to the model. For example, with some modification to the expressions for surface tension (see equation 4) and pressure source (see equation 16), the model can be directly used to assess the risk associated with formation of hydrogen blisters in aluminum fusion reactors [70], or indeed any metal / metal oxide pairs.

The present work focuses on anti-corrosion coatings, but is also applicable to osmotic blistering occurring on non-metallic or passive surfaces coated with other types of elastic coatings. For example, bituminous and polymer-modified bituminous roofings or coatings on concrete, as commonly observed on roads, bridges, buildings, and other infrastructures, often experience osmotic blistering [71]. For bituminous coatings on concrete, this can enhance corrosion of reinforcement bars affixed within [72]. In this case, a distinction needs to be made between waterproofed [73, 71] and non-waterproofed coatings [74]. The former does not allow water to pass through the coating; instead, water passes from the concrete structure underneath, which acts as an inelastic, semi-permeable membrane. This modifies the form of the transport equation, i.e., equation 10 considerably. Regardless of the direction from which water is transported into the blister, two further modifications are needed to adapt the present work for coated concrete: (1) the soluble components can arise from the concrete mixture [75, 76] and (2) the coating may exhibit orthotropic elasticity [73]. Nevertheless, the generality of the present work is such that we still expect the conclusions in section 4.1 to hold despite these mismatches in the exact statement on the model.

The current state of anti-corrosion and bituminous coating design incorporates a significant amount of empirical observation, as evidenced by the plethora of standardized tests for coating development published by the International Organization for Standardization (ISO), the National Association of Corrosion Engineers (NACE), etc. Some of these tests can be time- and resource-intensive (see [36, 71, 37, 38], among many others). Classification based on the dimensionless numbers introduced in 4.1 can provide an affordable theoretical alternative for coating development, although more experimental verification is needed.

Our model could also be applied in medical settings, e.g., formation of blisters on the skin upon exposure to heat sources [77, 78] or after excessive rubbing [79], the latter possibly exacerbated by hereditary illnesses, e.g., epidermolysis bullosa. In this context, the pressure source is the heart, from whence we make several predictions: (1) blistering should be more extensive when it occurs in the vicinity of arteries closer to the heart, (2) blistering should be more extensive in locations where the critical adhesion stress of neighboring pairs of skin layers is lower, (3) blisters exceeding a critical size undergo runaway growth - a situation which may assist in understanding the effect of extensive burns. We note that, in agreement with the prediction of the model, epidermolysis bullosa has been extensively linked to the inability to form functional structural proteins, which is expected to reduce critical adhesion stress [80, 81, 82]. The model is also more generally applicable to biological sciences, e.g., to understand the mechanics of biofilm growth [83] or deformation of dentin adhesives [84].

## 6. Conclusion

The present work builds upon existing understanding of blister formation to arrive at a comprehensive macroscopic model of osmotic blistering. Blisters initiate at locations with low critical adhesion stress and / or high local solute concentration and propagate by growth and / or deformation. Irreversible blister growth occurs when adhesion stress exceeds the critical adhesion stress of the metal / coating pair, and deformation occurs when osmotic pressure exceeds the elastic stress of the coating. If the growth continues unabated, the coating eventually becomes delaminated, and if deformation continues unabated, the coating eventually ruptures. The present work distinguishes between osmotic pressure arising from entrapped solutes and osmotic pressure arising from corrosion products. In the former, blister growth eventually stops due to decaying osmotic pressure; however, depending on the kinetics of delamination and the critical adhesion stress, the maximum elongation of the coating may still be exceeded, causing rupture. In the latter, we predict the existence of a critical delamination length, beyond which unstable blister growth occurs.

The model has been validated against blister growth data. We were able to confirm that blister growth is irreversibly driven by difference between adhesion stress and critical adhesion stress using two separate data sets employing polyurethane and alkyd coatings on steel. We were also able to predict the volume vs. time data of polyurethane coatings immersed in water with various amounts of undercoat magnesium acetate with no fitting parameters. We then use the model to develop three dimensionless numbers, each classified according to the source of osmotic pressure, which can be used for engineering design of polymeric coatings capable of resisting visible deformation, rupture, and delamination.

## Acknowledgement

We gratefully acknowledge support from Dow through the University Partnership Initiative and Colin Cwalina from Dow for providing guidance throughout this research project. We thank Dimitrios Fraggedakis and Hongbo Zhao for the insights they shared on this topic. We thank Ernie Latham-Brown and Yu Ren Zhou for the technical advice leading to the schematic in figure 6.

## 7. Supporting Information

### 7.1. Table of Nondimensionalized Equations

The nondimensional equations used in the present study is summarized in table 2.

| Equation | Note |
|---|---|
| $\overline{A} = \pi \overline{r}^2 \frac{2(1-\cos\psi)}{\sin^2\psi}$ | Stretched area |
| $\overline{V} = \frac{\pi \overline{r}^3}{3} \frac{\cos\psi\left(\cos^2\psi - 3\right) + 2}{\sin^3\psi}$ | Blister volume |
| $\bar{\pi} = \frac{2\bar{\gamma}\sin\psi}{\bar{r}}$ | See equation 11 |
| $\bar{\gamma} = \frac{\bar{E}}{6(1+K_{\mathrm{MR}})}\left[1 - \left(\frac{\sin^2\psi}{2[1-\cos\psi]}\right)^3\right]\left[1 - \frac{s_1}{s_{-1}}\frac{2(1-\cos\psi)}{\sin^2\psi}\right]$ | See equation 12 |
| $\frac{2(1-\cos\psi)}{\sin^2\psi} \leq 1 + \epsilon$ | See equation 13 |
| $\frac{d\bar{r}}{dt} = \overline{k}_{\mathrm{ad}}\left(\overline{\sigma}_{\mathrm{ad}} - \overline{\sigma}_{\mathrm{ad}}^*\right) H\left(\overline{\sigma}_{\mathrm{ad}} - \overline{\sigma}_{\mathrm{ad}}^*\right)$ | See equation 14 |
| $\overline{\sigma}_{\mathrm{ad}} = \bar{\gamma}\frac{2(1-\cos\psi)}{\sin^2\psi}$ | See equation 15 |
| $\bar{\pi} = \sum_i \frac{\bar{n}_i}{\overline{V}} - \overline{C}_{i,\mathrm{out}}$ | See equation 16 |
| $\frac{d\bar{n}_i}{dt} = \overline{J}_i \overline{A} + \nu_i \overline{R}_s \overline{A}_{\mathrm{base}} + \eta_i \overline{R}_p \overline{V}$ | See equation 17 |
| $\overline{J}_i = \delta_i\left(\overline{C}_{i,\mathrm{out}} - \overline{C}_i\right)\frac{2(1-\cos\psi)}{\sin^2\psi}$ | See equation 18 |
| $\overline{R}_s = -\frac{\delta_{O2}}{\nu_{O2}}\left(\overline{C}_{O2,\mathrm{out}} - \overline{C}_{O2}\right)\left[\frac{2(1-\cos\psi)}{\sin^2\psi}\right]^2$ | See equation 19 |
| $\overline{R}_p = \overline{k}_p\left(-\eta_{\mathrm{Fe}^{2+}}\ln\overline{C}_{\mathrm{Fe}^{2+}} - \eta_{\mathrm{OH}^-}\ln\overline{C}_{\mathrm{OH}^-} - \overline{\mu}^{\ominus}\right)$ | See equation 22 |
| $\bar{\mu}^{\ominus} = \ln\left[\left(-\frac{z_{OH^-}}{z_{Fe2+}}\right)^{\eta_{OH^-}}\frac{1 + \overline{K}_{FeCl+}\overline{C}_{Cl^-}}{1 + \overline{K}_{FeCl+}\overline{C}_{Cl^-,\mathrm{out}}}\right]$ $- \left(\frac{\overline{\beta}_p}{\overline{C}_{Fe(OH)2}}\right)^2$ | See equation 23 |

Table 2: Complete set of nondimensionalized equations used in the present work. Note that the conventional $s_1$ and $s_{-1}$ parameters used to describe Mooney-Rivlin solids (see equation 12) has been replaced with an equivalent pair of parameters: nondimensional Young's modulus $\overline{E}$ and stress equilibrium constant $K_{\mathrm{MR}}$.

### 7.2. Data Records

The data sets corresponding to figure 8(d) and figure 5(a) are recorded for posterity in table 3 and table 4, respectively.

| | | |
|---|---|---|
| 1.89 $\mu$g | 6.40 $\mu$g | 8.50 $\mu$g |

| t [days] | Vol. [nL] | t [days] | Vol. [nL] | t [days] | Vol. [nL] |
|---|---|---|---|---|---|
| 7.8 | 8.3 | 6.0 | 22.8 | 3.5 | 30.5 |
| 29.2 | 17.0 | 16.1 | 42.3 | 30.4 | 86.5 |
| 65.7 | 29.7 | 23.7 | 61.2 | 56.4 | 109.0 |
| 78.6 | 22.4 | 73.7 | 97.2 | 77.5 | 130.6 |
| 95.4 | 32.1 | 98.0 | 107.4 | 107.1 | 154.3 |
| 115.7 | 28.0 | 103.9 | 122.8 | 142.0 | 171.2 |
| 134.7 | 28.4 | 122.6 | 127.0 | 149.2 | 186.2 |
| 149.3 | 31.1 | 156.6 | 144.3 | | |

Table 3: Blister volume data for the undercoat salt immersion experiment described in [25]. Note that this data set is extracted from a figure published therein.

| Water mole fraction $x_{H_2O}$ | Water diffusivity in polyurethane $D_{H_2O}$ [$\mu m^2$/hr] |
|---|---|
| 0.65 | 37.5 |
| 0.85 | 70.5 |
| 0.90 | 89.9 |
| 0.93 | 108.1 |

Table 4: Diffusion data for the steady-state Payne cup measurement described in [25]. Note that this data set is extracted from a figure published therein.

### 7.3. Table of Experiments

The list of experiments performed to validate the critical delamination length hypothesis is shown in table 5. The result for experiment 9 has been shown in figure 9; in figure 11, we summarize other findings of interest.

The top row of figure 11 shows the replicate run of figure 9, with similar qualitative results albeit with earlier rust spot nucleation and faster blister growth. In particular, we still observe dark-green coloration at the edge of the blister at all points in time and the eventual development of pocket-shaped patterns. Differences in observation can be explained by the manual introduction of pinhole defects, which is difficult to control; otherwise, the observations made in section 4 are replicable.

The middle row of figure 11 shows the result of experiment 10, which is identical to experiment 9, except no pinhole defect is added. This delays rust spot nucleation, which first appears at 144 hours, although the growth of the blister proceeds at a typical speed once nucleation has begun. Once again, a dark-green patina is observed at the edge of the blister at all points in time. We note that the nucleation event occurs at the edge of the Teflon patch (dotted red line) and grows more slowly along the Teflon patch, suggesting that the Teflon patch imparts some degree of corrosion resistance. This justifies our hesitation

| Number | Pigmented? | Teflon? | Diameter [mm] | Pinhole defect? |
|:---:|:---:|:---:|:---:|:---:|
| 1 | Yes | Yes | ~5 | Yes |
| 2 | Yes | Yes | ~5 | No |
| 3 | Yes | Yes | ~15 | Yes |
| 4 | Yes | Yes | ~15 | No |
| 5 | Yes | No | - | Yes |
| 6 | Yes | No | - | No |
| 7 | No | Yes | ~5 | Yes |
| 8 | No | Yes | ~5 | No |
| 9 | No | Yes | ~15 | Yes |
| 10 | No | Yes | ~15 | No |
| 11 | No | No | - | Yes |
| 12 | No | No | - | No |

Table 5: List of critical delamination length experiments. Each experiment is repeated twice.

in claiming that the result in figure 9 supports the scaling relation $r \sim \exp t$.

The replicate run for the middle row of figure 11 does not show any undercoat rust, except the one occurring at the edge of the measurement area. More generally, all experiments on non-pigmented coatings with no pinhole defect do not develop undercoat rust, with the exception of the result presented in the middle row of figure 11. This confirms the hypothesis made by Funke that the addition of a pinhole defect can initiate osmotic blistering [23].

The bottom row of figure 11 shows the result for experiment 4, which is identical to experiment 10 except for the use of $TiO_2$ pigment particles. The frequency of blister formation is greatly increased relative to non-pigmented coatings. We originally suspected that $TiO_2$ particles either participate in or encourage the formation of pinhole defects around which blisters grow; however, this hypothesis conflicts with the observation that no unstable growth occurs, unlike that seen in the top and middle rows of figure 11. The lack of growth suggests that these blisters occur due to entrapped soluble components, implying that the presence of $TiO_2$ encourages the formation of pockets of soluble components. The exact mechanism is not clear; we speculate on adsorption of ions on the charged surface of $TiO_2$.

The second and third images of the bottom row of figure 11 are identical except for a dotted red line showing the location of the underlying Teflon patch. Comparing these two images, we observe that the edge of the Teflon patch exhibits a greatly increased frequency of blister formation. It is not clear why this should be the case; we speculate on the entrapment of soluble components at the edge of the Teflon patch, which could also explain the location of the nucleation event in the middle row of figure 11. The same observations are also made for the replicate run of experiment 4.

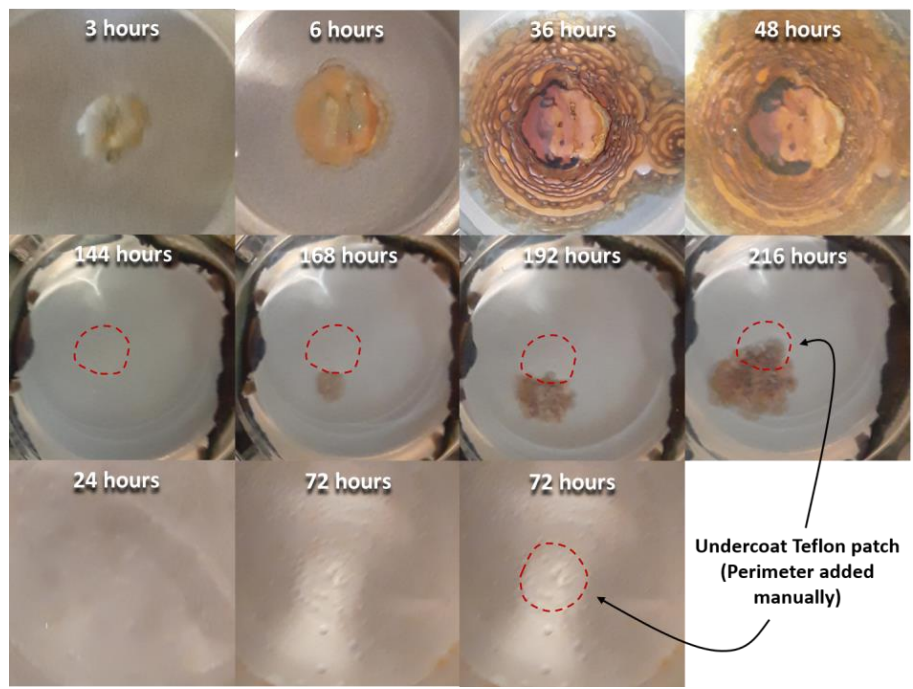

Figure 11: Top row: replicate result for experiment 9. Middle row: experiment 10. Bottom row: experiment 4.